\documentclass{article}

\usepackage{algorithm}
\usepackage{algorithmic}
\usepackage{amsmath}
\usepackage{amssymb}
\usepackage{graphics}
\usepackage{graphicx}

\newcommand{\pn}{{(\mathbb{R}^*_+)^n}}
\newcommand{\rn}{{\mathbb{R}^n}}
\newcommand{\sgn}{\text{sgn}}

\newtheorem{thm}{Theorem}[section]
\newtheorem{defi}{Definition}[section]
\newtheorem{prop}{Proposition}[section]
\newtheorem{lemme}{Lemma}[section]

\newenvironment{proof}
   {\noindent {\sc proof.} \\}
   {\begin{flushright}$\Box$\end{flushright}}

\begin{document}

\title{Approximation of Dynamical Systems using S-Systems Theory: 
		 Application to Biological Systems\footnote{This work is part of 
		 the CALCEL project, funded by "La Région Rh\^one-Alpes", France.}}

\author{Laurent Tournier}
\date{january 2005}

\maketitle

\begin{abstract}
In this article we propose a new symbolic-numeric algorithm to find positive 
equilibria of a $n$-dimensional dynamical system. This algorithm implies a 
symbolic manipulation of ODE in order to give a local approximation of 
differential equations with power-law dynamics (S-systems). A numerical 
calculus is then needed to converge towards an equilibrium, giving at the same 
time a S-system approximating the initial system around this equilibrium. This 
algorithm is applied to a real biological example in 14 dimensions which 
is a subsystem of a metabolic pathway in Arabidopsis Thaliana.
\end{abstract}

\section{Introduction}
The modelling and study of biological or biochemical systems has become an
exciting challenge in applied mathematics. The complexity of real biological 
dynamical systems lies essentially in the non-linearities of the dynamics as 
well as in the huge dimension of systems, often leading to a numerical approach.
However, as the understanding of cellular mechanisms grows, it has 
become obvious that the modelling step  strongly needs symbolic tools in order 
to manipulate more and more information and data, and to improve computational 
tools. Therefore a new area emerged, called "systems biology". It involves 
different fields of applied mathematics, from computer algebra (see for instance
\cite{laubenbacher}) 
to numerical computation (\cite{goldbeter}).\\
In the past decades, a lot of different frameworks have been developped to 
study behaviors of complex biochemical processes. Let us cite here three of 
them: the discrete networks (see the work of R. Thomas \cite{thomas}), the 
piecewise linear systems (the so-called Glass networks \cite{glass}, see also 
\cite{dejong}) and sigmoidal switch systems (\cite{plahte}). The main goal of 
all these 
approaches is to propose a (more or less) generic class of dynamical systems, 
either discrete or differential, that model some behaviors of complex 
interaction systems. Once this class is clearly defined, its mathematical 
relevance generally allows both theoretical and numerical analysis.\\
The class of systems we use in this paper is the set of S-systems (see 
\cite{mishra}, \cite{savvoi}, \cite{sav1}). The basic idea of this model is to 
represent interactions
between biochemical species with power-law dynamics. Their mathematical
expression is 
quite general, but sufficiently simple to allow theoretical and practical 
investigations. We propose in this article a symbolic-numeric algorithm that is
based upon S-systems theory. Its goal is to compute the positive equilibria of a
$n$-dimensional system of ordinary differential equations (ODE). As it converges
towards an equilibrium, it provides a S-system that approaches the original
dynamics around this equilibrium. As we will see, the local approximation of
some dynamics with power-laws can be made symbolically in any point of the phase
space. It can also include treatment of symbolic parameters. However, 
iterating this process in order to converge towards equilibria needs a
numerical computation, which prevents the use of pure symbolic tools to the
end.\\
In the following, we give a definition of the S-system class as it can be found
in the litterature(see for instance \cite{savvoi}). We then
propose a symbolic-numeric algorithm that computes an iteration leading to the
positive equilibria of a dynamical system. We will see an application of this
algorithm on a biological example in dimension $14$. We finally conclude with
some remarks on our algorithm and some future works.

\section{S-systems}

\subsection{Definition}
We give here a definition of the class of S-systems.
\begin{defi}
\label{defSS}
A $n$-dimensional S-system $S(\alpha,\beta,G,H)$ is a dynamical system defined 
by the $n$ differential equations:
$$
\dot{x_i} = \alpha_i \prod_{j=1}^{n} x_j^{g_{ij}} - 
							\beta_i \prod_{j=1}^{n} x_j^{h_{ij}} 
\; ,\;\; i = 1\dots n
$$
with 
$\alpha = \left( 
		\alpha_1, \dots, \alpha_n
	\right)\in \pn$
, $\beta = \left( 
		\beta_1 \dots \beta_n
	\right) \in \pn$\\ 
and $G = \left( g_{ij} \right)_{i,j=1\dots n} \in \mathcal{M}_n(\mathbb{R})$, 
$H = \left( h_{ij} \right)_{i,j=1\dots n} \in \mathcal{M}_n(\mathbb{R}).$\\
$\mathbb{R}^*_+$ denotes the set of strictly positive real numbers and
$\mathcal{M}_n(\mathbb{R})$ denotes the set of real square matrices of order
$n$.
\end{defi}
Let us introduce the vector field $F$ defined on $\Omega = \pn$:
$$
F(x) = \left(
\begin{array}{c}
f_1(x_1,\dots,x_n)\\
\vdots\\
f_n(x_1,\dots,x_n)
\end{array}
\right)
$$
with:
$$f_i(x_1\dots,x_n) =
\alpha_i \prod_{j=1}^{n} x_j^{g_{ij}} - \beta_i \prod_{j=1}^{n} 
x_j^{h_{ij}} \,,\quad i=1\dots n$$
$F$ is $\mathcal{C}^1$ and therefore locally lipschitz on the open $\Omega$.
Cauchy-Lipschitz theorem ensures the existence and unicity of a maximal 
solution of $S(\alpha,\beta,G,H)$ in $\Omega$, 
given any initial condition $x(0)=x^0\in\Omega$.

This definition of S-systems with power-law differential equations is strongly 
linked with equations of chemical kinetics. As an example, if we consider the 
following chemical pathway:
$$
A+2B \stackrel{k_1}{\rightarrow} C \stackrel{k_2}{\rightarrow} 3D+E
$$
then the mass-action law applied to species $C$ gives the equation:
$$
\frac{dc}{dt} = k_1ab^2 - k_2d^3e
$$
(capital letters design species and small letters design concentrations)\\
Therefore in definition \ref{defSS}, coefficients $\alpha_i$ and $\beta_i$ are 
sometimes called {\it kinetic rates} while $g_{ij}$ and $h_{ij}$ are called 
{\it kinetic orders}.

S-systems are part of a broader formalism known as 
quasi-monomial (QM) systems (see \cite{brenig}). An interesting result shows that QM systems can 
be expressed in the form of Lotka-Volterra quadratic systems (see \cite{brenig} 
for  details).

\subsection{Equilibrium points}
The study of the phase portrait of a S-system $S(\alpha,\beta,G,H)$ begins with
the search for equilibrium points in $\Omega$.
To find them, we have to solve the system:
\begin{equation}
\alpha_i\prod_{j=1}^{n} x_j^{g_{ij}} = \beta_i 
\prod_{j=1}^{n} x_j^{h_{ij}} \,, \quad i = 1\dots n
\label{eq_equi}
\end{equation}
In this paper we will use the following notation:\\
Given a vector $x\in \pn$ and a real square matrix $A = 
\left(a_{ij}\right)_{i,j=1\dots n}$, we
define the vector $x^A\in\pn$ by: 
$$(x^A)_i = \prod_{j=1}^{n} x_j^{a_{ij}}\,,\quad i = 1\dots n$$
With this notation, we can express equation (\ref{eq_equi}) as 
follows:
$$x^{G-H} = b$$
where $b$ is the vector $\left( \beta_1/\alpha_1, \dots, \beta_n/\alpha_n 
\right)$.\\
Taking the neperian logarithm, this equation leads to:
$$
(G-H) \ln x = \ln b
$$
(the logarithm is applied to all components of vector, i.e. $\ln x$ is the
$n$-dimensional vector $(\ln x_1,
\dots, \ln x_n)$). Posing $y=\ln x$, we are brought back to 
the resolution of a $n$-dimensional linear system in $y$.\\
We have therefore the following proposition:
\begin{prop}
A S-system $S(\alpha,\beta,G,H)$ has a unique equilibrium $\widetilde{x}$ in
$\Omega$ (i.e. a positive equilibrium) if and only if the matrix $(G-H)$ is
invertible. $\widetilde{x}$ can be calculated by the formula:
\begin{equation}
\widetilde{x} = b^{(G-H)^{-1}}
\label{eq_equilibre}
\end{equation}
\end{prop}

\subsection{Stability analysis of the equilibrium}

The stability analysis of the equilibrium $\widetilde{x}$ uses the study of the 
spectrum of $J_F(\widetilde{x})$ (the
jacobian of $F$ in $\widetilde{x}$). 
The question we tackle here is to find some relationship between the stability 
of $\widetilde{x}$ and some properties of the matrix $G-H$.\\
As a motivating example, let us consider the one-dimensional case. A
one-dimensional S-system is expressed by a single differential equation:
$$
S(\alpha,\beta,g,h)\,:\quad \dot{x} = f(x) = \alpha x^g - \beta x^h
$$
where $\alpha, \beta >0$ and $g,h\in \mathbb{R}$. The positive equilibrium
$\widetilde{x}$ of $(S)$ exists and is unique if and only if $g-h\neq0$. In this
case, an obvious calculation leads to:
$$
\frac{\partial f}{\partial x}(\widetilde{x}) = \alpha \widetilde{x}^{g-1} (g-h)
$$
so the stability of $\widetilde{x}$ depends directly on the sign of $g-h$: it 
is asymptotically stable if $g-h <0$ and unstable if $g-h>0$, regardless of
parameters $\alpha$ and $\beta$.

In the $n$-dimensional case, the stability depends on the sign of the real 
parts of the jacobian's eigenvalues. Derivating the functions 
$f_i(x_1\dots,x_n)$, we obtain, for $i,j=1\dots n$:
\begin{equation}
\label{eq_jacGH}
\frac{\partial f_i}{\partial x_j} (\widetilde{x}) =
\frac{\alpha_i}{\widetilde{x}_j} \prod_{k=1}^n \widetilde{x}_k^{g_{ik}} .\,
(g_{ij}-h_{ij})
\end{equation}
As in one-dimensional case, we thus obtain a formula that links the jacobian 
of $F$ in
$\widetilde{x}$ with the matrix $G-H$. However, it is not trivial to link 
the spectrum of $J_F(\widetilde{x})$ with the spectrum of $G-H$.\\
Let us recall here the definition of stability of matrices:
\begin{defi}
\label{defst}
A real square matrix $A$ of order $n$ is said to be stable (resp. semi-stable) 
if all its
eigenvalues $\lambda_i$, $i = 1\dots n$, have a negative (resp. non positive)
real part.
\end{defi}
We could hope that the stability of matrix $G-H$ was sufficient to deduce the
stability of $\widetilde{x}$. However this is not true, as we can see in the
following example.\\
For $n = 2$,  consider the S-system:
$$
(S1) : \left\{
\begin{array}{lll}
\dot{x} & = & 3 x y^2 - 2 x^4\\
\dot{y} & = & 4 x^3 y^4 - x^5 y^3
\end{array}
\right.
$$
we have:
$$
\alpha = \left( \begin{array}{l} 3\\ 4 \end{array} \right),\,
\beta = \left( \begin{array}{l} 2\\ 1 \end{array}\right),\,
G = \left( \begin{array}{ll} 1&2\\ 3&4 \end{array}\right),\,
H = \left( \begin{array}{ll} 4&0\\ 5&3 \end{array}\right)
$$
The matrix $G-H$ is equal to:
$$
G-H = \left( \begin{array}{ll} -3&2\\ -2&1 \end{array}\right)
$$
Its characteristic polynomial is $\chi(\lambda) =
(\lambda+1)^2$ so the matrix is stable.\\
Since $G-H$ is invertible, there is a unique equilibrium: 
$\widetilde{x} = \left( \frac{32}{3},\frac{256}{9} \right)$.We can calculate 
the two 
eigenvalues $\lambda_1$ and $\lambda_2$ of $J_F(\widetilde{x})$. 
We find that $\lambda_1,\lambda_2 > 0$, implying that 
$\widetilde{x}$ is an unstable node. As a result, in spite of the stability of 
matrix $G-H$, the equilibrium $\widetilde{x}$ is unstable.

The stability of $G-H$ is therefore insufficient to deduce the stability of
$\widetilde{x}$.\\
We need a stronger property known as sign stability
(see \cite{jefvdd},\cite{mayqui}).
\begin{defi}
Two real square matrices of order $n$, $A = \left(a_{ij}\right)_{i,j=1\dots n}$ 
and $B = \left(b_{ij}\right)_{i,j=1\dots n}$, have the same sign pattern if:
$$
\forall \, i,j = 1\dots n\,,\; \sgn(a_{ij}) = \sgn(b_{ij})
$$
The function $\sgn$ is the classical signum function:
$$
\forall x\in\mathbb{R}\,,\; \sgn(x) = \left\{
\begin{array}{cl}
+1 & \text{ if } x>0\\
0 & \text{ if } x=0\\
-1 & \text{ if } x<0\\
\end{array}
\right.
$$
\end{defi}
\begin{defi}
A real square matrix $A$ of order $n$ is said to be sign stable 
(resp. sign semi-stable) if 
all the matrices that have the same sign pattern
are stable (resp. semi-stable) in the sense of definition \ref{defst}.
\end{defi}
In \cite{jefvdd} we find a characterization of the sign
semi-stability:
\begin{thm}[Quirk-Ruppert-Maybee]
\label{th_qrm}\ \\
A real square matrix $A = \left(a_{ij}\right)_{i,j=1\dots n}$ is sign 
semi-stable if and only if it
satisfies the following three conditions:\\[2mm]{}
\begin{tabular}{l}
(i) $\forall\, i = 1\dots n\,,\; a_{ii} \leq 0$\tabularnewline[2mm]
(ii) $\forall\, i\neq j \,,\; a_{ij}a_{ji} \leq 0$\tabularnewline[2mm]
(iii) for each sequence of $k\geq 3$ distinct indices $i_1,\dots,i_k$,\\
$\qquad$we have: $ a_{i(1)i(2)}\dots a_{i(k-1)i(k)}a_{i(k)i(1)} = 0$
\end{tabular}
\\[2mm]{}
 (The third condition is equivalent to the fact that the directed graph
associated to $A$ admits no $k$-cycle for $k\geq3$)
\end{thm}
With this notion, we can formulate the following proposition, which links the
stability of the equilibrium $\widetilde{x}$ of a S-system with the sign
semi-stability of matrix $G-H$:
\begin{prop}
\label{prop_stabilite}
Let consider a $n$-dimensional S-system $S(\alpha,\beta,G,H)$. We assume that
$G-H$ is invertible and we note $\widetilde{x}$ the unique positive equilibrium
of $(S)$. We also assume that $\widetilde{x}$ is hyperbolic (i.e. none of the
eigenvalues of the jacobian of $F$ in $\widetilde{x}$ have zero real part).\\ If 
the matrix $G-H$ is sign semi-stable (i.e. if it verifies 
the three conditions of theorem \ref{th_qrm}) then, regardless of parameters 
$\alpha$ and $\beta$, the equilibrium $\widetilde{x}$ is
asymptotically stable.
\end{prop}
\begin{proof}
Let us note $J$ the Jacobian of $F$ in $\widetilde{x}$ and $P$ the matrix $G-H$.
The equation \ref{eq_jacGH} yields:
$$
\frac{\partial f_i}{\partial x_j} (\widetilde{x}) =
\frac{\gamma_i}{\widetilde{x}_j} p_{ij}
$$
with $\gamma_i = \alpha_i \prod_{k=1}^n \widetilde{x}_k^{g_{ik}}$. As
$\gamma_i>0$ and $\widetilde{x}_j>0$ for all $i$ and $j$, matrices $J$ and $P$
have the same sign pattern. We can thus deduce that $J$ is semi-stable and as
$\widetilde{x}$ is supposed hyperbolic, it is asymptotically stable.
\end{proof}
Let us remark that the latter equation gives, in
matricial notation:
$$
J = \Gamma P D^{-1}
$$
where $\Gamma$ and $D$ are diagonal matrices:
$$
\Gamma=\left(\begin{array}{ccc}\gamma_1&&\\&\ddots&\\&&\gamma_n\end{array}
\right)\,, \qquad D=\left(
\begin{array}{ccc}\widetilde{x}_1&&\\&\ddots&\\&&\widetilde{x}_n
\end{array}\right)
$$
so $\sgn(\det(J)) = \sgn(\det(P))$. As we have supposed that $P$ is invertible,
we deduce that $J$ is also invertible and does not have null eigenvalues. 
So we supposed the hyperbolicity of $\widetilde{x}$ in order to avoid 
imaginary eigenvalues of $J$.\\
We can easily verify in the previous example that $G-H$ is stable but not sign
semi-stable ($g_{22}-h_{22} = 1 >0$).

\section{Local approximation of dynamical system using S-systems}

In this part, we propose an algorithm for approaching the equilibria of a
dynamical system using S-systems. Simultaneously, we obtain a
S-system that approximates the initial system around the equilibrium.

\subsection{Monomial approximation of a positive vector field}
(see \cite{sav1},\cite{sav2},\cite{savvoi}).\\
Let's consider the positive vector field $F : \pn \rightarrow \pn$.
$$
F(x) = 	\left( \begin{array}{c}
	f_1(x_1,\dots,x_n)\\
	\vdots\\
	f_n(x_1,\dots,x_n)
			\end{array} \right)
$$
We will suppose $F$ sufficiently smooth on $\pn$.\\
Let us define the following change of variables: $y = \ln x$, and 
express the logarithm of $F(x)$ as a function $G$ of the new variable 
$y$: 
$$
\ln F(x) = \ln F(e^y) = G(y)
$$
The function G is sufficiently smooth on $\rn$. Given any arbitrary point 
$y^0\in \rn$, let
us write the Taylor expansion of $g_i$ (for $i=1\dots n$) in the neighborhood 
of $y^0$ at the first order: 
$$
\forall i = 1\dots n\,,\; g_i(y) = g_i(y^0) + \sum_{j=1}^n (y_j - y_j^0)
\frac{\partial g_i}{\partial y_j} (y^0) + o(\parallel y-y^0 \parallel)
$$
We introduce the functions $\tilde{g}_i(y)$ for $i=1\dots n$:
$$\forall i=1\dots n\,,\;\tilde{g}_i(y) = g_i(y^0) + \sum_{j=1}^n 
(y_j - y_j^0) \frac{\partial g_i}{\partial y_j} (y^0)$$
and the functions $\tilde{f}_i = \exp (\tilde{g}_i(y))$: 
$$
\begin{array}{lll}
\tilde{f}_i(x) &\stackrel{def}{=}& e^{\tilde{g_i}(y)}\\
& = & e^{g_i(y^0)} \exp \left( \displaystyle \sum_{j=1}^n (y_j-y_j^0)
      \frac{\partial g_i}{\partial y_j} (y^0) \right)\\
& = & e^{g_i(y^0)} \displaystyle \prod_{j=1}^n \exp \left( (y_j-y_j^0)
      \frac{\partial g_i}{\partial y_j} (y^0) \right)
\end{array}
$$
As $y=\ln x$ and $g_i(y)=\ln f_i(x)$, we have:
$$
\tilde{f}_i(x) = f_i(x^0) \prod_{j=1}^n \left( \frac{x_j}{x_j^0} 
\right)^{\frac{\partial g_i}{\partial y_j} (y^0)}
$$
and:
$$
\begin{array}{lll}
\displaystyle
\frac{\partial g_i}{\partial y_j} (y) &=& \displaystyle
   \frac{\partial}{\partial y_j} \left( \ln (f_i(e^y)) \right)\\[3mm]{}
&=& \displaystyle \frac{1}{f_i(e^y)} \frac{\partial}{\partial y_j} 
    \left(  f_i(e^y) \right)\\[3mm]{}
&=& \displaystyle \frac{1}{f_i(x)} e^{y_j} \frac{\partial f_i}{\partial x_j} 
    ( e^y )\\[3mm]{}
&=& \displaystyle \frac{x_j}{f_i(x)} \frac{\partial f_i}{\partial x_j} (x)\\
\end{array}
$$
Therefore, we have defined a vector field $\widetilde{F} = \displaystyle
\left(\tilde{f}_i\right)_{i=1\dots n}$
\begin{equation}
\label{ftilde}
\widetilde{F}(x) = \left( \alpha_i \prod_{j=1}^n x_j^{g_{ij}} \right)_{i=1\dots n}
\end{equation}
\begin{equation}
\label{coeffFtilde}
\text{with: }  \left\{
\begin{array}{lcl}
\alpha_i(x^0) &=& \displaystyle f_i(x^0) \prod_{j=1}^n (x_j^0)^{-g_{ij}}\\[2mm]{}
g_{ij}(x^0) &=& \displaystyle \frac{x_j^0}{f_i(x^0)} 
				\frac{\partial f_i}{\partial x_j}(x^0)
\end{array}
\right.
\end{equation}
The basic idea is to use the monomial vector field $\widetilde{F}$ as an
approximation of $F$ in a neighborhood of $x^0$.
\begin{defi}
\label{defSapp}
Let $F$ be a smooth $n$-dimensional vector field, $F:\pn\rightarrow\pn$ and 
$x^0$ any vector of $\pn$. We call S-approximation of $F$ in $x^0$ the
vector field $\widetilde{F}$ defined by equations (\ref{ftilde}) and
(\ref{coeffFtilde}).
\end{defi}

The following proposition is basic for what follows:
\begin{prop}
\label{propstab}
Let $F$ be a positive vector field and
$\widetilde{F}$  its S-approximation in $x^0$. The following 
equalities hold:
\begin{itemize}
\item $\widetilde{F}(x^0)=F(x^0)$
\item $\displaystyle \forall\, i,j=1\dots n\,,\; \frac{\partial 
\tilde{f}_i}{\partial x_j}
(x^0) = \frac{\partial f_i}{\partial x_j}(x^0)$\\
 (or, which is equivalent: $J_F(x^0) = J_{\widetilde{F}}(x^0)$)
\end{itemize}
\end{prop}

\subsection{Finding equilibria of a dynamical system}

We consider a $n$-dimensional dynamical system of the form:
$$
(S)\quad \dot{x} = V^+(x) - V^-(x)
$$
where $x$ lies in $\pn$ and $V^+$, $V^-$ are  positive vector fields.
$V^+,V^- : \pn\rightarrow\pn$. For $i=1\dots n$,
the term $v_i^+(x)$ is the {\em production term} of the variable $x_i$ and 
$v_i^-(x)$
the {\em decay term} of $x_i$. We propose an algorithm for finding 
an equilibrium point of $(S)$ that lies in $\pn$. Meanwhile, we get a
S-system that approximates the system $(S)$ around this equilibrium.

Given a point $x^0$ in $\pn$, we introduce the fields $\widetilde{V}^+$ and 
$\widetilde{V}^-$ which are the S-approximations of the
fields $V^+$ and $V^-$ in $x^0$. Let us consider the $n$-dimensional S-system: 
$$
(S_{x^0})\quad \dot{x} = \widetilde{V}^+(x) - \widetilde{V}^-(x)
$$
using (\ref{ftilde}) and (\ref{coeffFtilde}), we obtain:
$$
(S_{x^0}) :\quad \dot{x}_i =  \alpha_i \prod_{j=1}^{n} x_j^{g_{ij}} - 
							\beta_i \prod_{j=1}^{n} x_j^{h_{ij}} 
							,\quad i = 1\dots n
$$
where:
\begin{equation}
\label{eq_coeffab}
\left\{
\begin{array}{lll}
\alpha_i & = & \displaystyle v_i^+(x^0)\prod_{j=1}^n (x_j^0)^{-g_{ij}}\\[2mm]{}
\beta_i & = & \displaystyle v_i^-(x^0)\prod_{j=1}^n (x_j^0)^{-h_{ij}}\\
\end{array}
\right.
\end{equation}
and:
\begin{equation}
\label{eq_coeffgh}
\left\{
\begin{array}{lll}
	g_{ij} &=& \displaystyle \frac{x_j^0}{v_i^+(x^0)}
	\frac{\partial v_i^+}{\partial x_j}(x^0)\\[2mm]{}
	h_{ij} &=& \displaystyle \frac{x_j^0}{v_i^-(x^0)}
	\frac{\partial v_i^-}{\partial x_j}(x^0)
\end{array}
\right.
\end{equation}

If the matrix $G-H$ is invertible, the system $(S_{x^0})$ admits a
unique equilibrium $x_{eq} \in \pn$: 
$$
x_{eq} = b^{(G-H)^{-1}}
$$
with $b = \left( \beta_1/\alpha_1,\dots,\beta_n/\alpha_n
\right)$. This point $x_{eq}$ depends on the initial point $x^0$ where we made 
our approximation. Let $x^1 = x_{eq}$ be the new initial point where we make our
new S-approximation. The 
algorithm (\ref{alg1}) computes the iteration of that process.

\begin{algorithm}[ht]
\caption{Search of an equilibrium point of system $(S)$}
\label{alg1}
\begin{algorithmic}
\REQUIRE 
$$\begin{array}{ll}
X = x^0&\in\pn \text{ (initial condition)}\\ V^+,V^- &\text{: positive vector 
fields defined over }\pn\\ \epsilon >0 &\text{: precision }
\end{array}$$
\ENSURE
unless we fall in a degenerate case, we find a point $y$ close to a positive
equilibrium of $(S)$ with the precision $\epsilon$. Meanwhile, we obtain the
S-system $(S_y)$ that approximate system $(S)$ around this equilibrium.
\STATE 
\REPEAT
\STATE $Y := X$
\FOR{$i=1$ to $n$} 
\FOR{$j=1$ to $n$}
\STATE $g_{ij} := \displaystyle \frac{X_j}{v_i^+(X)} \frac{\partial
v_i^+}{\partial X_j}$
\STATE $h_{ij} := \displaystyle \frac{X_j}{v_i^-(X)} \frac{\partial
v_i^-}{\partial X_j}$
\ENDFOR
\STATE $\alpha_i := \displaystyle v_i^+(X)\prod_{j=1}^n
			(X_j)^{-g_{ij}}$
\STATE $\beta_i := \displaystyle v_i^-(X)\prod_{j=1}^n
			(X_j)^{-h_{ij}}$
\STATE $b_i := \beta_i / \alpha_i$
\ENDFOR
\IF{$\det(G-H)\neq0$}
\STATE $X := b^{(G-H)^{-1}}$ 
\ELSE 
\STATE degenerate case: algorithm terminated
$\rightarrow$ restart the algorithm with a new initial condition
\ENDIF
\UNTIL $\parallel X - Y \parallel < \epsilon$
\STATE Result $:= X$
\end{algorithmic}
\end{algorithm}

\subsection{Correctness of the algorithm}
Let's describe the first iteration.\\
Let $x^0\in\pn$. With formulae (\ref{eq_coeffab}) and (\ref{eq_coeffgh}), we 
define the quantities
$\alpha_i(x^0)$, $\beta_i(x^0)$, $g_{ij}(x^0)$ and $h_{ij}(x^0)$.They depend 
on the choice of the initial point $x^0$. We assume that the constructed 
matrices $G(x^0)$ and $H(x^0)$ verify the condition: $\det(G-H)\neq0$.
Thanks to this assumption, there exists a unique equilibrium point of the system
$(S_{x^0})$. We will denote it $x^1$, and we define the function $\Psi : \pn
\rightarrow \pn$ that, to each $x^0\in \pn$ associates the point $x^1$.\\ 
Our algorithm is iterative, in the sense that it computes: 
$$
(I) \left\{
\begin{array}{l}
x^0 \in \pn\\
x^{n+1} = \Psi (x^n)
\end{array}
\right.
$$
This iterative process converges towards fixed points of $\Psi$. However we do 
not {\it a priori} know if
all fixed points of $\Psi$ are indeed limits of $(I)$. In other words, we must 
find
which fixed points are {\em attracting}.\\ The correctness of the
algorithm (\ref{alg1}) is a consequence of the two following lemmas:
\begin{lemme}
 The equilibria of initial system $(S)$ are the fixed points of the
function $\Psi$
\end{lemme} 
\begin{lemme}
Given a fixed point $\bar{x}$ of $\Psi$, there exists some initial points $x^0$ that
lead to $\bar{x}$ by the iteration $(I)$. In other words, the positive equilibria of
$(S)$ are the attracting fixed points of $\Psi$.
\end{lemme} 
\begin{proof} (First lemma)
Let $\bar{x}\in\pn$ such that $\det(G(\bar{x})-H(\bar{x}))$ is different from 
zero.
(for convenience, we will omit the dependency in $\bar{x}$, and note for instance  
$G$ in place of $G(\bar{x})$). Using equation (\ref{eq_equilibre}), we have:
\begin{equation}
\label{eq_psi}
\Psi(\bar{x}) = b^{(G-H)^{-1}} 
\end{equation}
where $b$ is the vector $\left( \beta_i / \alpha_i \right)_{i=1\dots n}$.\\
Therefore: 
$$
\begin{array}{lll}
\Psi(\bar{x}) = \bar{x} & \Longleftrightarrow & b^{(G-H)^{-1}} = \bar{x}\\
 & \Longleftrightarrow & b = \bar{x}^{(G-H)}\\
 & \Longleftrightarrow & \displaystyle \forall i = 1\dots n ,\, 
   \frac{\beta_i}{\alpha_i} = \prod_{j=1}^n \bar{x}_j^{g_{ij}-h_{ij}}\\
 & \Longleftrightarrow & \displaystyle \forall i = 1\dots n ,\, 
   \beta_i \prod_{j=1}^n \bar{x}_j^{h_{ij}} = \alpha_i \prod_{j=1}^n 
	\bar{x}_j^{g_{ij}}\\
\end{array}
$$
By definition, $\alpha_i \prod_{j=1}^n \bar{x}_j^{g_{ij}}$ (resp. 
$\beta_i \prod_{j=1}^n \bar{x}_j^{h_{ij}}$) is the S-approximation of $V^+$
(resp. $V^-$) in $\bar{x}$. Proposition \ref{propstab} implies then:
$$
\Psi(\bar{x})  =  \bar{x} \quad \Longleftrightarrow \quad V^+(\bar{x}) = 
V^-(\bar{x})
$$ 
Thus, the equilibria of $(S)$ are the fixed points of the function $\Psi$.
\end{proof}

In order to prove the second lemma, we will use the following fixed point
criterion:\\
\begin{em}
If the function $\Psi$ is a contraction on the open set $W$ and if $\bar{x}\in 
W$ is a fixed point of $\Psi$, then $\bar{x}$ is the unique fixed point of 
$\Psi$ in $W$ and it is attracting, that is to say,
for all $x^0\in W$, the iteration $(I)$ converges towards $\bar{x}$.
\end{em}
\begin{proof} (Second lemma)
Let $\bar{x}$ be a fixed point of $\Psi$. We assume that
$\det(G(\bar{x})-H(\bar{x}))\neq0$. The continuity of the determinant implies
that there exists a neighboorhood $W$ of $\bar{x}$ in which $\det(G-H)\neq0$. 
To prove that $\bar{x}$ is attracting, it is sufficient to show that $\Psi$ is 
contracting in a neighboorhood of $\bar{x}$. For that, we show that the 
jacobian of $\Psi$ in $\bar{x}$ is zero.\\
Using (\ref{eq_coeffab}) and (\ref{eq_coeffgh}) and posing:
$$
\left\{
\begin{array}{lll}
U^+ &=& \log (V^+)\\
U^- &=& \log (V^-)\\
U &=& U^+ - U^- = \displaystyle \log \left( \frac{V^+}{V^-} \right)\\
P &=& G-H
\end{array}
\right.
$$
we obtain, for all $x\in W$: 
$$
\Psi_i(x) = \prod_{j=1}^n \left( \frac{v_j^+(x)}{v_j^-(x)} 
\right)^{p_{ij}^{(-1)}(x)}
$$
where $\displaystyle \left( p_{ij}^{(-1)} \right)_{i,j=1\dots n}$ is the inverse
of the matrix $P = G - H$.\\
Let's calculate $p_{ij}^{(-1)}$:
$$
\begin{array}{lll}
p_{ij} &=& g_{ij}-h_{ij}\\
 &=& \displaystyle x_j \left(\frac{1}{v_i^+} 
 \frac{\partial v_i^+}{\partial x_j}
  - \frac{1}{v_i^-} \frac{\partial v_i^-}{\partial x_j} \right)\\[3mm]{}
&=& \displaystyle x_j \left( \frac{\partial u_i^+}{\partial x_j}
  - \frac{\partial u_i^-}{\partial x_j} \right)\\[3mm]{}
&=& x_j \displaystyle \frac{\partial u_i}{\partial x_j}
\end{array}
$$
in matricial notation: $P = J_u(x) \Delta\,$ where $J_u(x)$ is the 
jacobian of the function $U$ evaluated in $x$ and $\Delta$ is the 
diagonal matrix:
$$
\Delta = \left( 
\begin{array}{lll}
x_1 & &\\
& \ddots &\\
& & x_n
\end{array}
\right)
$$
Therefore $P^{-1} = \Delta^{-1} \left( J_u(x)\right)^{-1} = \Delta^{-1} 
\left( J_{u^{-1}}(x)\right)$ ($u^{-1}$ is the reciprocal function of $u$) 
and so:
$$
\forall i,j=1\dots n\,,\quad
p_{ij}^{(-1)} = \frac{1}{x_i} \frac{\partial u_i^{-1}}{\partial x_j}
$$
with (\ref{eq_psi}) we have, for $i=1\dots n$ and $x\in W$:
$$
\begin{array}{lll}
\Psi_i(x) &=& \displaystyle x_i \prod_{j=1}^n \exp\left( -\frac{1}{x_i} u_j(x) 
\frac{\partial u_i^{-1}}{\partial x_j}(x) \right)\\
&=& \displaystyle x_i \exp\left( - \sum_{j=1}^n \frac{u_j(x)}{x_i}  
\frac{\partial u_i^{-1}}{\partial x_j}(x) \right)
\end{array}
$$
Deriving this (and omitting the dependency in $x$), we get, for $k\neq i$:
$$
\frac{\partial \Psi_i}{\partial x_k} = \left[ \displaystyle \sum_{j=1}^n u_j 
\frac{\partial^2 u_i^{-1}}{\partial x_j \partial x_k} \right]
\exp\left( - \sum_{j=1}^n \frac{u_j(x)}{x_i}  
\frac{\partial u_i^{-1}}{\partial x_j} \right)
$$
and

$\displaystyle
\frac{\partial \Psi_i}{\partial x_i} = \left[ \displaystyle \sum_{j=1}^n u_j 
\frac{\partial^2 u_i^{-1}}{\partial x_j \partial x_k}  - \frac{1}{x_i}
\sum_{j=1}^n u_j \frac{\partial u_i^{-1}}{\partial x_j}\right] \times
$
\vspace{-0.6cm}
\begin{flushright}
$\displaystyle
\exp\left( - \sum_{j=1}^n \frac{u_j(x)}{x_i}  
\frac{\partial u_i^{-1}}{\partial x_j} \right)
$
\end{flushright}
As we have shown that the fixed points of $\Psi$ are the equilibria of $(S)$, we
deduce that $\forall k = 1\dots n,\, u_k(\bar{x}) = 0$, therefore: 
$$
J_{\Psi}(\bar{x}) = 0
$$
We deduce that $\Psi$ is contracting in a neighboorhood of $\bar{x}$, and then
that $\bar{x}$ is attracting. This concludes the proof of the second lemma and 
the correctness of the algorithm.
\end{proof}

\subsection{An example with multiple positive equilibria}

\begin{figure}[htbp]
\begin{center}
\includegraphics[scale=0.5]{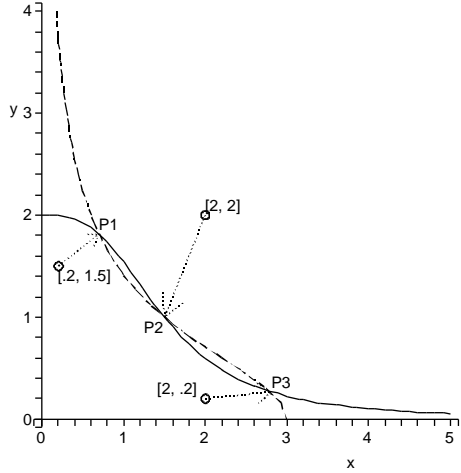}
\caption{nullclines of system (\ref{eq_switch}). (The dash line represents 
$f_1(x,y)=0$ and the solid one represents $f_2(x,y)=0$). 
The central equilibrium $P2$ can be
shown to be unstable while the two others, $P1$ and $P3$ are stable. The arrows
represent the three experimentations described in the text.}
\label{fig_switch}
\end{center}
\end{figure}

We present here the application of our algorithm for a dynamical system having
multiple positive equilibrium points. It is a system known as {\em biological 
switch} (see \cite{switch}).\\
Let's consider the two dimensional dynamical system:
\begin{equation}
\label{eq_switch}
\left\{
\begin{array}{lll}
\dot{x} &=& \displaystyle \frac{3}{1+y^2} - x\\[3mm]{}
\dot{y} &=& \displaystyle \frac{6.75}{3.375+x^3} - y
\end{array}
\right.
\end{equation}
It represents the temporal evolution of two positive quantities $x$ and $y$ with
linear decay and sigmoidal production (we use here the Hill function
$\displaystyle H^-(z) = \frac{K^n}{K^n+z^n}$ often used by biologists to model
sigmoidal interactions). As we can see on figure
\ref{fig_switch}, This system shows three equilibrium points. The
values of these points can be calculated:
$$
P1\approx \left( \begin{array}{l} 0.697\\1.818 \end{array} \right) \quad 
P2= \left( \begin{array}{l} 1.5\\1.0 \end{array} \right) \quad 
P3\approx \left( \begin{array}{l} 2.802\\0.266 \end{array} \right)  
$$
We can show that $P2$ is unstable whereas $P1$ and $P3$ are stable (cf.
\cite{switch}).

Applying our program in {\sf Maple}, we found three different initial conditions
each of which tends towards one of the three equilibrium points (see figure
\ref{fig_switch} and numerical results below). The convergence appears
to be fast since we need only 4 iterations to approach the equilibria with a
precision of $10^{-5}$. We will discuss about the convergence speed in part
\ref{cv}.

\begin{itemize}
\item With initial condition $x^0 = (2,2)$, algorithm finished in 4
iterations and found $P2$ with a precision of $10^{-5}$. The numerical 
S-system obtained is given by:
$$
\left\{
\begin{array}{lll}
\dot{x} &=& 1.500\, y^{-1} - x\\
\dot{y} &=& 1.837\, x^{-1.5} -y
\end{array}
\right.
$$
\item With initial condition $x^0 = (0.2,1.5)$, algorithm finished in 4
iterations and found $P1$ with a precision of $10^{-5}$. The numerical 
S-system obtained is given by:
$$
\left\{
\begin{array}{lll}
\dot{x} &=& 1.745\, y^{-1.535} - x\\
\dot{y} &=& 1.647\, x^{-0.274} - y
\end{array}
\right.
$$
\item With initial condition $x^0 = (2,0.2)$, algorithm finished in 4
iterations and found $P3$ with a precision of $10^{-5}$. The numerical 
S-system obtained is given by:
$$
\left\{
\begin{array}{lll}
\dot{x} &=& 2.352\, y^{-0.132} - x\\
\dot{y} &=& 3.879\, x^{-2.6} - y
\end{array}
\right.
$$
\end{itemize}

\subsection{Stability analysis of approximate S-systems}
Consider the $n$-dimensional dynamical system:
\begin{equation}
\label{sysdyn}
\left\{
\begin{array}{l}
\displaystyle \frac{dx}{dt} = F(x) = V^+(x) - V^-(x)\\
x \in \pn
\end{array}
\right.
\end{equation}
Algorithm \ref{alg1} ensures that, given any initial condition $x^0$ in $\pn$,
unless we fall in a degenerate case, we produce a sequence
$(x^q)_{q\in\mathbb{N}}$ (with $x^q\in\pn$) that tends towards a limit  point 
$\widetilde{x}\in\pn$ which is an equilibrium of (\ref{sysdyn}). More
precisely, $x^q = \Psi^{(q)}(x^0)$.\\
Meanwhile, at each step, it provides us with a S-system $S_q(\alpha_q, \beta_q,
G_q, H_q)$ which comes from the S-approximations of functions $V^+$ and
$V^-$ in $x^q$. Thus, we have:
$$
\left\{
\begin{array}{lll}
\alpha_q &=& \alpha(x^q)\\
\beta_q &=& \beta(x^q)\\
G_q &=& \displaystyle \left(g_{ij}^q\right)_{i,j=1\dots n} 
     \text{ with: }g_{ij}^q = g_{ij}(x^q) \\
H_q &=& \displaystyle \left(h_{ij}^q\right)_{i,j=1\dots n} 
     \text{ with: }h_{ij}^q = h_{ij}(x^q) \\
\end{array}
\right.
$$
where $\alpha$, $\beta$, $g_{ij}$ and $h_{ij}$ are the functions defined in
(\ref{coeffFtilde}). If we assume that $V^+$ and $V^-$ are at least
$\mathcal{C}^1$, we deduce that these sequences converge, as $q$ tends to
$\infty$, towards:
$$
\left\{
\begin{array}{lllll}
\alpha_q &\rightarrow & \alpha(\widetilde{x}) 
		&\stackrel{def}{=}&\widetilde{\alpha}\\
\beta_q &\rightarrow & \beta(\widetilde{x}) 
		&\stackrel{def}{=}&\widetilde{\beta}\\
G_q &\rightarrow & G(\widetilde{x}) 
		&\stackrel{def}{=}&\widetilde{G}\\
H_q &\rightarrow & H(\widetilde{x}) 
		&\stackrel{def}{=}&\widetilde{H}\\
\end{array}
\right.
$$
Let $(\widetilde{S})$ be the following S-system:
\begin{equation}
\label{sysapp}
(\widetilde{S})\,:\; \dot{x_i} = \widetilde{\alpha}_i 
\prod_{j=1}^{n} x_j^{\tilde{g}_{ij}} - 
\widetilde{\beta}_i \prod_{j=1}^{n} x_j^{\tilde{h}_{ij}},
\; i = 1\dots n
\end{equation}
We want to know in what sense the system (\ref{sysapp}) approach the system
(\ref{sysdyn}). 
An answer is given by the following proposition:
\begin{prop}
$F$ is supposed $\mathcal{C}^r$ ($r\geq 1$). The equilibrium $\widetilde{x}$ of 
(\ref{sysdyn}) is an equilibrium of (\ref{sysapp}). Moreover, 
if $\widetilde{x}$ is hyperbolic, then the flow generated by 
(\ref{sysapp}) is topologically conjugate to the flow generated by 
(\ref{sysdyn}) in a neighborhood of~$\widetilde{x}$.
\end{prop}
\begin{proof}
The first assertion is obvious with proposition \ref{propstab}. Let prove the
second assertion: it is a direct consequence of the Hartman-Grobman theorem
(see for instance \cite{wiggins}).\\ 
Proposition \ref{propstab} shows that systems  (\ref{sysdyn}) and 
(\ref{sysapp}) have the same linearized dynamical systems in $\widetilde{x}$.
Thanks to the Hartman-Grobman theorem, we know that these systems are 
topologically conjugate to their linearized dynamical systems. By transitivity
of the topological conjugation, we deduce that (\ref{sysdyn}) and (\ref{sysapp})
are topologically conjugate around $\widetilde{x}$.
\end{proof}
This proposition implies that the stability of $\widetilde{x}$ for system 
(\ref{sysapp}) is the same that the stability of $\widetilde{x}$ for system 
(\ref{sysdyn}). As an exemple, let us consider the following 2-dimensional 
dynamical system:
$$
(Ex)\left\{
\begin{array}{lll}
\dot{x} &=& \displaystyle \frac{x}{2+y} - \frac{x^2y^4}{(3+x)(4+y^3)}\\[3mm]{}
\dot{y} &=& \displaystyle \frac{5x}{3+x} - \frac{2xy^3}{(x+1)(y+2)}\\
\end{array}
\right.
$$
We find the equilibrium point $\widetilde{x}\approx(1.2301,1.6950)$ and the
matrix $\widetilde{G}-\widetilde{H}$:
$$
\widetilde{G}-\widetilde{H} = \left(
\begin{array}{cc}
-0.709&-2.812\\
0.261&-2.541
\end{array}
\right)
$$
Thanks to theorem \ref{th_qrm}, we see that $\widetilde{G}-\widetilde{H}$ is
sign semi-stable. The point $\widetilde{x}$, as equilibrium of
$(Ex)$ is hence stable.

\section{Application to a biological example}

We present here a current work we are doing in collaboration with 
G. Curien
(see \cite{curien}). The goal of this work is to understand 
the metabolic system responsible for the  
synthesis of aminoacids in Arabidopsis Thaliana. 
So far, we have focused our study on a subsystem of $14$ variables, with
$9$ symbolic parameters. The
differential equations present several strongly nonlinear terms due to
allosteric control of some enzymes~; in particular, Hill functions and
compositions of Hill functions. Since the latter are rational 
functions, seeking positive equilibria is equivalent to solving a polynomial
system. Algebraic manipulations have led us to a simplified system with $5$
polynomial equations in $5$ variables. Because of the complexity of these
equations, we were not able to achieve the resolution of this system with 
purely symbolic computations and manipulating symbolic parameters (we used 
{\sf Maple} 8 and 9).
That is why symbolic-numeric methods appeared as a satisfactory way to tackle
this problem. As it is a system of equations coming from biochemical kinetics,
S-systems seemed to be an appropriate tool in this work.\\
{\it In vivo}, this system exhibits a stationnary behavior.
Giving realistic values of parameters, we managed, thanks to our algorithm, to
find this positive equilibrium. We now have to study the S-approximation of the
system near this equilibrium, with different realistic sets of parameters. An
interesting idea is also to propose a piecewise S-approximation of the system in
order to reproduce its behavior in a wider zone of the phase space.
This work is in progress.

\section{Discussions and concluding remarks}

\subsection{Convergence of our algorithm\label{cv}}
\begin{figure}[htbp]
\begin{center}
\includegraphics[scale=0.5]{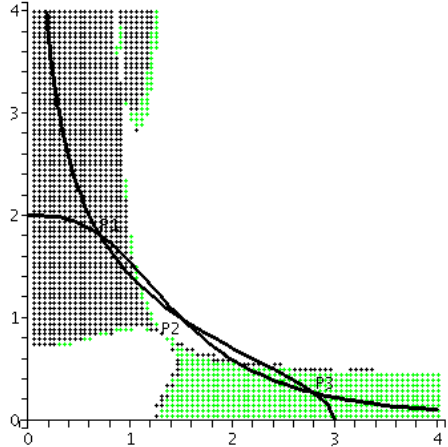}
\caption{Basins of attraction of points $P1$ (dark), $P2$ (white) and $P3$ 
(grey). We obtained these graphs by applying algorithm \ref{alg1} for system 
(\ref{eq_switch}) with initial conditions taken in a regular grid of $]0,4]$.}
\label{bassins}
\end{center}
\end{figure}

The algorithm described above computes the iterations of a vectorial function 
$\Psi$ on an initial point $x^0\in\pn$, in order to converge towards a fixed
point of $\Psi$. As the jacobian of $\Psi$ is the null matrix in those fixed
points, we know that the convergence speed is very fast (up to four or five
iterations in all the examples presented, for a precision of $10^{-4}$ or 
$10^{-5}$). As a matter of fact, we are in a case where the speed of convergence
is the best possible. Indeed, if the function $\Psi$ is $K$-contractant, one 
can easily verify that the convergence of the iteration 
is in $K^n$ (where $n$ is the number of iterations). Since 
$J_{\Psi}(\widetilde{x})=0$, then we can find a neighborhood 
of $\widetilde{x}$ wherein $\Psi$ is $K$-contractant for any $0<K<1$.\\
However, even if the speed of convergence is very fast, the algorithm
behaviour is strongly dependent on the choice of initial point $x^0$. Indeed, if
initial system has multiple positive equilibria, each of them have distinct
basins of attraction. We cannot {\it a priori} know in which of these basins is
the point $x^0$. We even cannot ensure that $x^0$ actually lye in one of them.
In fact, the study of basins of attractions of such iterations is a complex
issue. The boundaries of such basins can be quite complicated, even fractals 
\cite{fractal}. As an example, we launched our
algorithm for the switch system (equations (\ref{eq_switch})) with initial 
conditions taken on a grid of $]0,4]^2$. To vizualize the three basins, we 
colored the initial points (fig \ref{bassins}).

\subsection{interaction between symbolic and numerical calculus}

As we said in the introduction, a large part of research concerning the 
analysis of biological phenomena uses both symbolic and numerical techniques. 
The S-systems as we described represent a large class of systems, yet their 
simple mathematical expression allows symbolic manipulations, providing 
a practical framework of study. Algorithm~\ref{alg1}, as presented here needs
numerical estimations of symbolic parameters. Nevertheless the technique of 
S-approximation (def 
\ref{defSapp}) consists of symbolic manipulations (in particular, we use 
symbolic computation of partial derivatives). It can be calculated in any point 
of the phase space and can include symbolic parameters.\\
S-approximation gives a computable and rather good approximation of ODE systems 
(see \cite{sav1} for a comparison between power-law approximation and 
linearization). A very interesting idea is therefore to use the context 
information (given for instance by biologists) of a particular system in order 
to create a piecewise S-approximation of this system. This should provide a 
global approximation interpolating the system in some critical points in the 
phase space (see \cite{sav4}).

\bibliographystyle{plain}
\bibliography{maBiblio}  

\begin{thebibliography}{10}

\bibitem{mishra}
M.~Antoniotti, A.~Policriti, N.~Ugel, and B.~Mishra.
\newblock Xs-systems : extended s-systems and algebraic differential automata
  for modeling cellular behavior.
\newblock {\em Proceedings of the International Conference on Hih Performance
  Computing, HiPC 2002}, pages 431--442, 2002.

\bibitem{brenig}
L.~Brenig and A.~Goriely.
\newblock Universal canonical forms for time continuous dynamical systems.
\newblock {\em Phys. Rev. A}, 40:4119--4121, 1989.

\bibitem{switch}
J.L. Cherry and F.R. Adler.
\newblock How to make a biological switch.
\newblock {\em J. Theor. Biol.}, 203:117--133, 2000.

\bibitem{curien}
G.~Curien, S.~Ravanel, and R.~Dumas.
\newblock A kinetic model of the branch-point between the methionine and
  threonine biosynthesis pathways in arabidopsis thaliana.
\newblock {\em Eur. J. Biochem.}, 270(23):4615--4627, 2003.

\bibitem{dejong}
H.~de~Jong, J.-L. Gouzé, C.~Hernandez, M.~Page, S.~Tewfik, and J.~Geiselmann.
\newblock Qualitative simulation of genetic regulatory networks using
  piecewise-linear model.
\newblock {\em Bull. Math. Biol.}, 66(2):301--340, 2004.

\bibitem{glass}
L.~Glass.
\newblock Combinatorial and topological methods is nonlinear chemical kinetics.
\newblock {\em J. Chem. Phys.}, 63, 1975.

\bibitem{goldbeter}
A.~Goldbeter.
\newblock {\em Biochemical oscillations and cellular rhythms}.
\newblock Cambridge University Press, 1996.

\bibitem{fractal}
C.~Grebogi and E.~Ott.
\newblock Fractal basin boundaries, long-lived chaotic transients, and
  unstable-unstable pair bifurcation.
\newblock {\em Phys. Rev. Lett.}, 50(13):935--938, 1983.

\bibitem{jefvdd}
C.~Jeffries, V.~Klee, and P.~Van~Den Driessche.
\newblock When is a matrix sign stable ?
\newblock {\em Can. J. Math.}, 29(2):315--326, 1976.

\bibitem{laubenbacher}
R.~Laubenbacher.
\newblock A computer algebra approach to biological systems.
\newblock {\em Proceedings of the 2003 International Symposium on Symbolic and
  Algebraic Computation (ISSAC)}, 2003.

\bibitem{mayqui}
J.~Maybee and J.~Quirk.
\newblock Qualitative problems in matrix theory.
\newblock {\em SIAM Review}, 11(1):30--51, 1969.

\bibitem{plahte}
T.~Mestl, E.~Plahte, and S.W. Omholt.
\newblock A mathematical framework for describing and analyzing gene regulatory
  networks.
\newblock {\em J. Theor. Biol.}, 176:291--300, 1995.

\bibitem{sav4}
M.A. Savageau.
\newblock Alternative designs for a genetic switch: analysis of switching times
  using the piecewise power-law representation.
\newblock {\em Math. Biosci.}, 180:237--253, 2002.

\bibitem{sav2}
M.A. Savageau and E.O. Voit.
\newblock Recasting nonlinear differential equations as s-systems~: a canonical
  nonlinear form.
\newblock {\em Math. Biosci.}, 87:83--115, 1987.

\bibitem{thomas}
R.~Thomas and M.~Kaufman.
\newblock Multistationarity, the basis of cell differentiation and memory. i.
  structural conditions of multistationarity and other non-trivial behaviour,
  and ii. logical analysis of regulatory networks in terms of feedback
  circuits.
\newblock {\em Chaos}, 11:170--195, 2001.

\bibitem{savvoi}
E.O. Voit.
\newblock {\em Computational analysis of biochemical systems}.
\newblock Cambridge University Press, 2000.

\bibitem{sav1}
E.O. Voit and M.A. Savageau.
\newblock Accuracy of alternative representations for integrated biochemical
  systems.
\newblock {\em Biochemistry}, 26:6869--6880, 1987.

\bibitem{wiggins}
S.~Wiggins.
\newblock {\em Introduction to applied nonlinear dynamical systems and chaos}.
\newblock Springer Verlag, 1990.

\end{thebibliography}
\end{document}